\documentclass[sigconf, balance=false]{acmart} 
\usepackage{tabularx}
\usepackage{multirow}

\AtBeginDocument{%
  \providecommand\BibTeX{{%
    \normalfont B\kern-0.5em{\scshape i\kern-0.25em b}\kern-0.8em\TeX}}}

\setcopyright{acmcopyright}
\copyrightyear{2023}
\acmYear{2023}
\acmDOI{10.1145/3544548.3580780}

\acmConference[CHI '23]{2023 CHI Conference on Human Factors in Computing Systems}{April 23-28,
  2023}{Hamburg}
\acmPrice{15.00}
\acmISBN{978-1-4503-XXXX-X/18/06}

\begin{document}


\title[QButterfly: Lightweight Survey Extension for Online User Interaction Studies]{QButterfly: Lightweight Survey Extension for Online User Interaction Studies for Non-Tech-Savvy Researchers}

\author{Nico Ebert}
\email{nico.ebert@zhaw.ch}
\orcid{0000-0002-9683-4792}
\affiliation{%
  \institution{Zurich University of Applied Sciences, School of Management and Law}
  \streetaddress{Theaterstrasse 17}
  \city{Winterthur}
  \state{Zurich}
  \country{Switzerland}
  \postcode{8401}
}

\author{Björn Scheppler}
\affiliation{%
  \institution{Zurich University of Applied Sciences, School of Management and Law}
  \streetaddress{Theaterstrasse 17}
  \city{Winterthur}
  \state{Zurich}
  \country{Switzerland}
  \postcode{8401}
}

\author{Kurt Ackermann}
\affiliation{%
  \institution{Zurich University of Applied Sciences, School of Management and Law}
  \streetaddress{Theaterstrasse 17}
  \city{Winterthur}
  \state{Zurich}
  \country{Switzerland}
  \postcode{8401}
}

\author{Tim Geppert}
\affiliation{%
  \institution{Zurich University of Applied Sciences, School of Management and Law}
  \streetaddress{Theaterstrasse 17}
  \city{Winterthur}
  \state{Zurich}
  \country{Switzerland}
  \postcode{8401}
}

\renewcommand{\shortauthors}{Ebert et al.}

\begin{abstract}
We provide a user-friendly, flexible, and lightweight open-source HCI toolkit (\href{http://github.com/QButterfly}{github.com/QButterfly}) that allows non-tech-savvy researchers to conduct online user interaction studies using the widespread Qualtrics and LimeSurvey platforms. These platforms already provide rich functionality (e.g., for experiments or usability tests) and therefore lend themselves to an extension to display stimulus web pages and record clickstreams. The toolkit consists of a survey template with embedded JavaScript, a JavaScript library embedded in the HTML web pages, and scripts to analyze the collected data. No special programming skills are required to set up a study or match survey data and user interaction data after data collection. We empirically validated the software in a laboratory and a field study. We conclude that this extension, even in its preliminary version, has the potential to make online user interaction studies (e.g., with crowdsourced participants) accessible to a broader range of researchers.
\end{abstract}



\maketitle

\section{Introduction}
\begin{figure}[b]
  \centering
  \includegraphics[width=1\columnwidth]{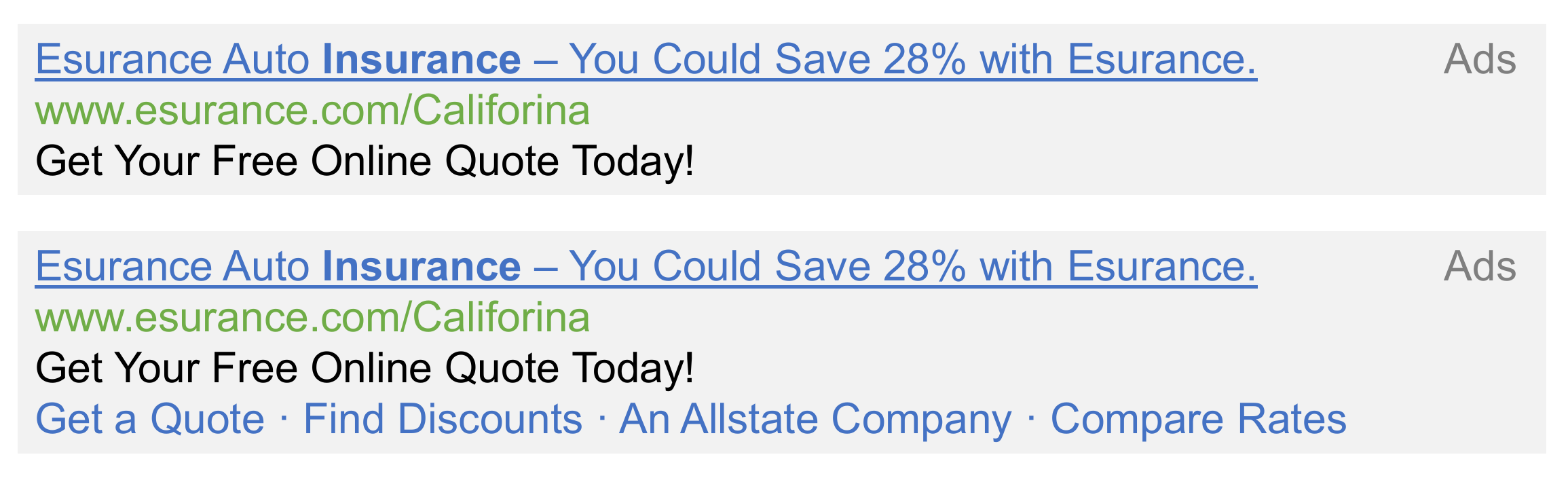}
  \Description{The figure shows two different bing ads, one in a control condition and a slightly different one in a treatment condition}
  \caption{Bing ads without (top) and with site links (bottom) as example stimuli in an online user interaction study \cite{kohavi_online_2017}}
  \label{fig:bing_ad}
\end{figure}

Companies such as Alphabet, Amazon, and Microsoft, permanently expose their users to user interaction studies in the form of simultaneous experiments to optimize their websites’ effectiveness \cite{kohavi_online_2007, tang_overlapping_2010}. For example, Microsoft’s Bing search engine team studied the effect of additional site links in their search engine ads (Figure \ref{fig:bing_ad}) and, as a result, was able to increase revenue by “tens of millions of dollars per year with neutral user impact” \cite{kohavi_online_2017}. These experiments are often carried out directly in the field with a specific percentage of (often unknowing) users involved over some time \cite{kohavi_online_2007}, and in-situ, as participants use their everyday devices instead of laboratory equipment. This makes it possible to observe user behavior in a real-world environment instead of an artificial one \cite{baravalle_remote_2003}. Platforms like Optimizely allow website owners to conduct online user interaction experiments on their websites in real-time \cite{optimizely_unlock_nodate}, and online user interaction studies do not necessarily have to be carried out in the form of experiments (e.g., randomization, different stimuli). Website owners can also ask users for qualitative feedback and integrate analytics tools, such as Matomo, Open Web Analytics, or Google Analytics, to observe user behavior and optimize their website accordingly (e.g., \cite{hasan_using_2009}). 

However, unlike in these mostly correlational studies conducted by practitioners, researchers in HCI and related disciplines typically want to develop and test a theory or design theory-guided artifacts. Therefore, they need a higher level of control than the abovementioned examples and different instruments to conduct studies (e.g., for screening, controlling for extraneous variables, or carrying out attention and manipulation checks). There is a rising demand among HCI researchers to conduct controlled studies at scale \cite{egelman2014crowdsourcing, findlater2017}. Therefore, participants are often recruited from a panel or platform, such as Amazon Mechanical Turk (MTurk), and directed to other platforms that help to conduct surveys, experiments, or usability tests (e.g., \cite{kittur_crowdsourcing_2008}).

Despite the popularity of controlled online user interaction studies, researchers have many technical challenges to overcome \cite{hirth2017crowdsourcing}. First, it remains laborious for scientists to conduct controlled online user interaction studies, especially with a large number of participants (e.g., collecting data from multiple tools that present either survey or stimulus elements) \cite{hirth2017crowdsourcing}. Second, researchers who wish to develop such studies need to have software development skills and be able to develop the required functions in a customized manner (e.g., to identify participants during the study) \cite{hirth2017crowdsourcing}. Lastly, researchers and study participants need to use multiple and often isolated tools and websites (e.g., survey website, stimulus website), which is especially error-prone in a complex study design with many users \cite{hirth2017crowdsourcing}. We are not aware of any easy-to-use and open-source software suitable for settings that (i) allow the use of standard experimental features, such as randomization, presentation of multiple stimuli, and flexible questions and, at the same time, (ii) ensures a consistent stimulus presentation even when users complete the study using their own private equipment.

To address this need, this paper introduces the open-source QButterfly HCI toolkit, which allows non-tech-savvy users to design controlled user interaction studies using the widely available Qualtrics and LimeSurvey platforms. QButterly has already successfully been used in HCI research (e.g., \cite{ebert_bolder_2021, ebert2023saliency}). Our software enhances existing survey platforms with stimulus presentation and user-tracking features and consists of (i) a premade survey template containing embedded JavaScript to manage the website presentation and collect user-tracking data, (ii) a small JavaScript library to be embedded into the website that tracks the user data and reports them to the survey platform, and (iii) Excel scripts based on generic regular expressions to analyze the collected data. We present two validation studies demonstrating consistent data collection in a laboratory and a realistic, crowdsourced setting.

\section{Related Work}
The provision of toolkits that expand the existing repertoire of researchers falls into the categories of “constructive research” \cite{oulasvirta2016} and “code as a contribution” \cite{fogarty2017code}. There are several examples where such toolkits have contributed significantly to scientific research in the HCI domain and beyond (e.g., AWARE \cite{ferreira2015}, psiTurk \cite{gureckis2016psiturk}, oTree \cite{chen2016otree}). To benefit other researchers, toolkits must be evaluated with adequate measures (e.g., demonstration, usage, or a technical benchmark) \cite{ledo2018evaluation}.

Researchers who want to conduct controlled user interaction studies typically employ surveys, experiments, or usability tests \cite{lazar2017research} and need a collection of tools to present a stimulus (e.g., multiple websites), collect data from users (e.g., via an online questionnaire) and observe user behavior (e.g., via electronic notes). Stimuli can range from purely visual design ideas with no user interaction to clickable website prototypes with limited user interaction or fully developed websites with multiple interlinked web pages. In the context of websites, to capture user interaction, the user must be monitored along a “visitor path” across multiple linked web pages \cite{jansen_understanding_2009, hirth2017crowdsourcing}. A relatively simple tracking technique involves capturing mouse clicks on specific areas of the stimulus (e.g., links clicked on a website) or dwell time (e.g., time spent on a website before visiting another one) \cite{jansen_understanding_2009}. More sophisticated tracking techniques may involve recording mouse trajectories, keyboard inputs, or eye movements (e.g., to generate visual heat maps) \cite{acar_no_2020, nielsen_eyetracking_2010}.

Whereas in traditional controlled user interaction studies in the lab (e.g., with students using lab computers), researchers have a high level of control and can interact with the participants, controlled \textit{online} user interaction studies with crowdsourced subjects using their own personal devices pose different challenges \cite{findlater2017}. Especially in studies with many anonymous participants, researchers have limited abilities to assist participants (e.g., to prevent input mistakes). Researchers could easily handle procedures in the lab, such as assigning a user to an experimental condition, conducting an ex-ante survey (e.g., screening), asking a user to visit several condition-specific websites, and conducting an ex-post survey (e.g., usability experience). However, in online settings, this would not only be prone to error (e.g., users accidentally switching between survey and stimulus windows on their device) but require software development (e.g., keeping track of a specific user’s actions). To create studies with a better user experience and a lower error susceptibility at a large scale, researchers would have to overcome the technical disintegration of the required tools \cite{hirth2017crowdsourcing}. They would have to develop software that neatly integrates surveying, stimulus presentation, and behavior monitoring - at the price of reinventing functionalities provided out of the box by existing web analytics and survey tools.

Such requirements for controlled online user interaction studies also apply to other types of online studies. For example, some cognitive-psychological experiments, such as the Stroop test \cite{stroop1935studies}, require the precise measurement of stimulus presentation and reaction times. Many tools exist that support researchers in conducting these kinds of studies online and provide required timing precision \cite{bridges_timing_2020}. Some require programming skills, while others are even able to be used by non-tech-savvy researchers (e.g., QRTEngine \cite{barnhoorn_qrtengine_2015}, Gorilla \cite{anwyl-irvine_gorilla_2020}). However, the focus of their stimulus presentation is not on a realistic “look and feel” of a website (e.g., on a mobile device) as in user interaction studies, nor on a comprehensive assessment of user behavior. Therefore, these tools can neither present the required stimuli nor record the user interaction with them. 

To conduct controlled online user interaction studies, one could also consider versatile software frameworks for online studies such as psiTurk \cite{gureckis2016psiturk} or oTree \cite{chen2016otree}; however, researchers would need to have skills to develop the user interaction study in a programming language such as Python. For HCI researchers that have previously only used survey tools (e.g., Qualtrics) and have little to no programming skills or do not have the ressources to invest time and effort into software development, there is a hurdle to using these tools.

\section{Introducing QButterfly}
QButterfly provides a lightweight approach for controlled online user interaction studies that requires no special programming skills; it is published under the MIT license. Its preliminary version allows embedding HTML pages in Qualtrics and LimeSurvey, recording user clicks on the HTML elements, and directly storing them on the survey platforms. Thanks to the integration, a) existing features of the survey platforms can be used (e.g., question types, random assignment to conditions), and participants do not have to switch between different windows, b) it is not necessary to manually match survey data and click data after the study (e.g., via IP address as in \cite{goldstein_novel_2019}) and c) user click stream data can be analyzed in real-time during the survey (e.g., to influence the survey flow based on user clicks). For example, user ID matching would be required if a survey platform is combined with a web analytics suite, such as Matomo, to analyze clickstreams on the stimulus website. While such a combination may seem more powerful because advanced analytics features would be available in such cases, it is problematic because matching user IDs across platforms requires technical skills or is associated with a disadvantage for study participants (e.g., they need to enter an identifier on the stimulus website that needs to be stored). Using web analytics results to influence an individual survey flow in real-time during a survey would be even more complex to achieve for non-technical users.

\begin{figure}[t]
  \centering
  \includegraphics[width=1\columnwidth]{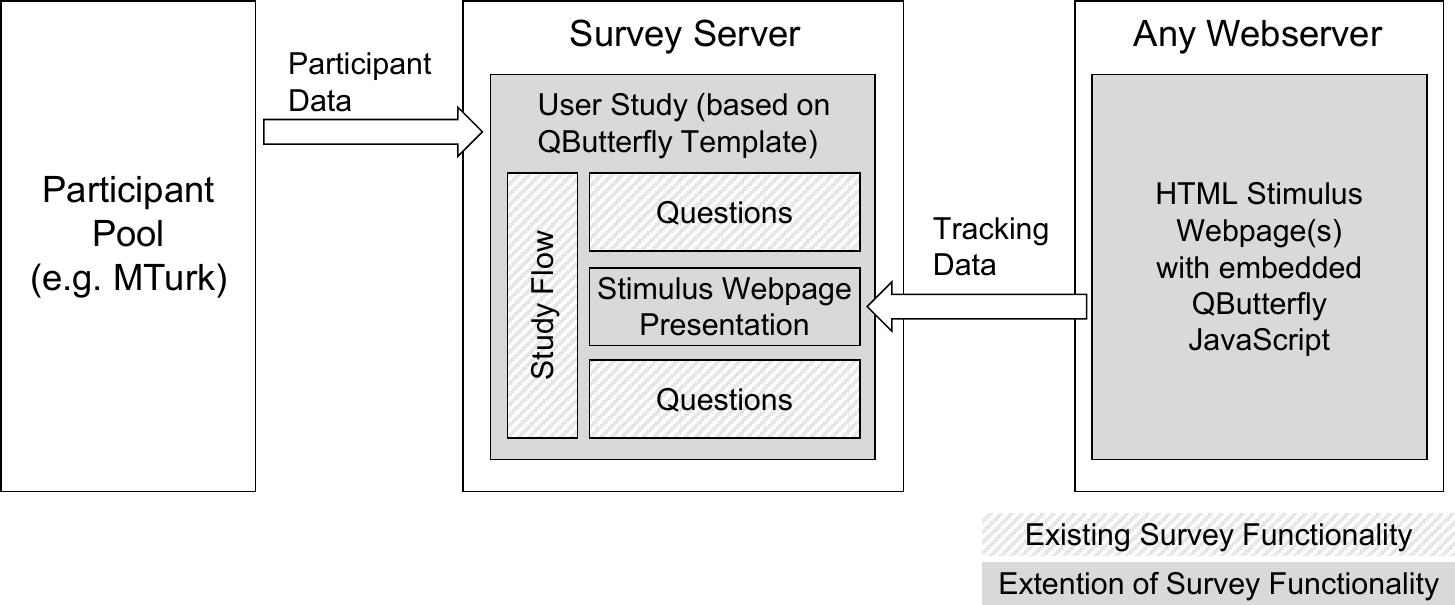}
  \caption{QButterfly architecture}
  \label{fig:architecture}
  \Description{The figure shows from left to right: a participant pool, the survey server, and the web server hosting the stimulus website. QButterfly uses the QButterfly template and controls the study flow via the survey. It displays the website within the survey.}
\end{figure}

Figure \ref{fig:architecture} shows the system architecture of QButterfly, while users can be recruited from any participant pool (e.g., Prolific, MTurk). The QButterfly template hosted in a Qualtrics or LimeSurvey environment is used to set up online user studies (e.g., collect qualitative feedback or conduct experiments). Features, such as the random assignment of participants to different conditions and the design of questions, take advantage of the easy-to-use survey platform interface. The QButterfly survey template contains HTML and JavaScript code that allows the user to display the stimulus webpage(s) seamlessly in the survey. This can be as straightforward as one or multiple static HTML web pages without server backend functionality. Furthermore, each webpage is enriched with the QButterfly JavaScript library, which records the user clicks and transfers them to the survey server. That means access to the website is required, and user-interaction cannot be recorded on any Internet website.

\begin{figure}[t]
  \centering
  \includegraphics[width=0.7\columnwidth]{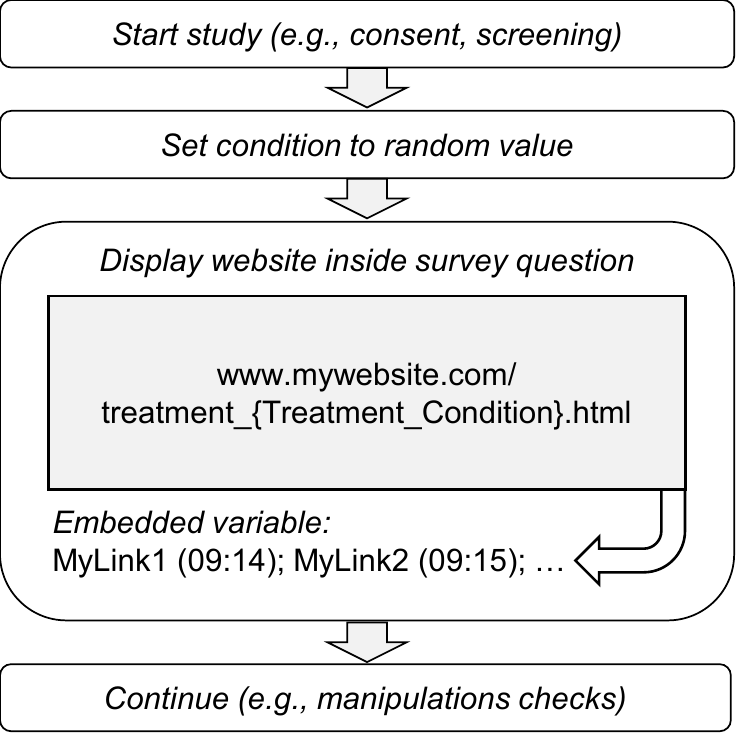}
  \caption{Illustrative survey flow on the survey platform with QButterfly}
  \label{fig:flow}
  \Description{A survey is started, participants are assigned to a condition, the condition-specific website is shown, tracking data is collected, and the survey is continued e.g., with manipulation checks}
\end{figure}

Figure \ref{fig:flow} illustrates an exemplary study flow with a simple between-subjects design in Qualtrics using QButterfly. At the study’s beginning, users consent to participate and are screened. Next, they are randomly assigned to an experimental condition that contains a particular survey question element. This unique question has three purposes: (i) to display the website, (ii) to “listen” to user events related to the website, and (iii) to record user events on the survey platform.

First, the question element can present a different website in each condition. Technically, it contains the HTML code of an inline frame (iframe), a broadly-supported HTML element (\cite{htmliframe}). An iframe is used to visually embed one browser window (“child”) within another browser window (“parent”). In the case of QButterfly, the stimulus website is embedded as a child window within the survey’s parent window. During interaction with the website, the surrounding Qualtrics elements (e.g., the button to go backward or forward) can be disabled to avoid an unwanted interruption in the website presentation. The termination of a specific user interaction episode can either be triggered by a user event (i.e., a click on a particular element), or a pre-defined timer that automatically activates survey elements or leads the participant to subsequent survey questions (e.g., manipulation checks, demographics).
Second, the survey question element contains JavaScript code that “listens” to events in the child window. These events are generated by the QButterfly JavaScript embedded in the stimulus website. The current version of the JavaScript code can record clicks on HTML elements, such as a hyperlink, together with a timestamp. 

Third, the questions record the user events in an “embedded field,” which can be accessed during the survey from within the survey environment. For example, data stored in the embedded field can be used to change the original survey flow or other questions. Also, the embedded data is stored in the regular survey dataset as additional variables and is thus readily accessible for data analysis.

\section{Creating and Analyzing a user interaction Study}
Our open-source software provides a user-friendly data collection and analysis pipeline for creating controlled online user interaction studies. A demonstration and all survey and code files are available online, including detailed instructions on setting up the software (\href{http://github.com/QButterfly}{github.com/QButterfly}). First, researchers import the “Qualtrics template” or “LimeSurvey template”, which implements a simple user study including a stimulus website. They can conduct the research and observe the recorded data through the survey platform. The template also records a user’s browser properties (e.g., type, version, screen resolution) by default, which are necessary parameters for pre-studies before entering the field. For example, it might be required to check the properties of a user’s device (e.g., screen dimensions) for screen-out purposes or to adapt the stimulus website accordingly. The following section describes the procedure for using QButterfly based on the “Qualtrics template”. The procedure for LimeSurvey is very similar and described in the GitHub repository. 

First, the embedded fields shown in Table \ref{tab:fields} must be configured on the survey platform. The only exception is the “eventStream” field, which collects click stream data. Most importantly, the address of the stimulus website(s) needs to be defined. Afterward, the QButterfly JavaScript must be embedded into each webpage of the website on which user behavior is tracked.

\begin{table*}[t]
  \caption{Embedded fields on the survey platform}
  \label{tab:fields}
  \begin{tabularx}{\textwidth}{lX}
    \toprule
    Field&Description\\
    \midrule
    windowURL & Address of stimulus website (e.g., https://www.mywebsite.com/index.HTML) \\
    windowBorder & Border size of stimulus website within survey window in pixels (e.g., “0” = no visible border) \\
    windowHeight & Height size of stimulus website within survey window in pixels (e.g., “640”) \\
    windowWidth & Width of stimulus website within survey window in pixels (e.g., “480”) \\
    windowScroll & Visible scrollbars of stimulus website within survey window (“yes” or “no”) \\
    eventStream & Chronological stream of tracking events of a user in the format “timestamp in ms since 01.01.1970 00:00:00 UTC\#Event\_ID” (e.g., 1630841029899\#ready\_demo.html; 1630841029900\#load\_demo.html; 1630841031050\#MyLink;) \\
  \bottomrule
   \multicolumn{2}{X}{Note: The standard fields browser type, browser version, operating system, screen resolution, java support, and user agent will also be collected by QButterfly}
\end{tabularx} 
\end{table*}

\begin{table*}[t]
  \caption{Excel functions for the analysis of the event stream}
  \label{tab:excel}
  \begin{tabularx}{\textwidth}{lX}
    \toprule
    Function&Description\\
    \midrule
    countEvent(Cell, Event\_ID)&Return the number of occurrences of a specific user event (e.g., MyLink) in an event stream. \\
    countEventPattern(Cell, Event\_ID\_1, …)&Returns the number of occurrences of a specific sequence of user events (e.g., MyLink1, MyLink2) in an event stream. \\
    timestamp(Cell, Event\_ID, occurrence)&Returns the timestamp (ms since 01.01.1970 00:00:00 UTC) of the n-th occurrence of a specific event in an event stream. \\
  \bottomrule
\end{tabularx} 
\end{table*}

\begin{figure*}[t]
  \caption{Code of an illustrative HTML page with a tracking script. The script is embedded in the header (1) and contains the URL of the survey server (2). The page contains an example link to be tracked, which is identified via the ID "MyLink" (3).}
  \centering
  \includegraphics[scale=1.2]{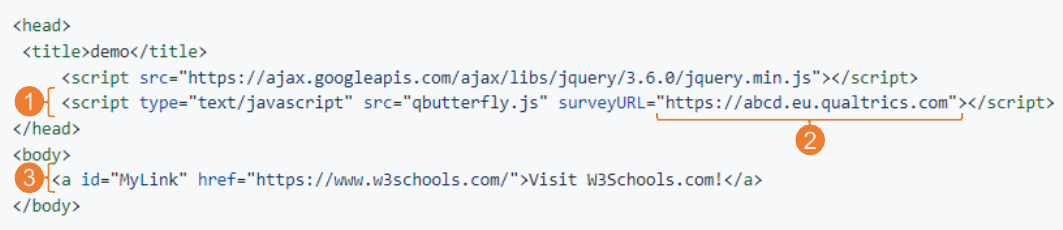}
  \Description{An html snippet is shown that shows how the QButterfly script is embedded. It can be found on the GitHub site.}
 \label{fig:html1}
\end{figure*}

Figure \ref{fig:html1} shows how the QButterfly JavaScript is embedded in the code of a simple webpage. To begin with, the required library JQuery is embedded (“jquery.min.js”). Next, “qbutterfly.js” is embedded as this contains the address of the survey server as a parameter (e.g., https://abcd.qualtrics.com). This allows the stimulus websites to send events to the survey server. Finally, the HTML element to be tracked (e.g., an image, link, checkbox) is marked with an ID tag with a unique name (e.g., MyLink). This is a simple hyperlink in Figure \ref{fig:html1}. This ID later identifies the specific user action within the events recorded in the “eventStream” field.

Events of a webpage are recorded in the “eventStream” field. Each recorded event consists of a timestamp (ms since 01.01.1970 00:00:00 UTC) and an event ID (e.g., 1629802674592\#MyLink). Events are separated by the character “;”. An event with the webpage name is generated when the first elements of a webpage begin to load, and the user starts to see the webpage (…\#ready\_demo.html)\footnote{QButterfly relies on the JQuery library to record all events. The first QButterfly event is triggered when the page’s Document Object Model (DOM) is ready to manipulate. This JQuery event is comparable to the DOMContentLoaded provided by most browsers. The second event is generated when all page resources, including all images, are loaded. This event listens to “window.onload” (for details, see https://api.jquery.com/ready/).}. A second event is generated after all static page elements (e.g., images, CSS, scripts) have been successfully loaded (…\#load\_demo.html). Additionally, each click on an element – whether previously defined with an ID or not – will generate an event (e.g., 1629802676308\#MyLink, 1629802677000\#Undefined). 

\begin{figure}[t]
  \caption{Code to hide HTML page until all static elements are loaded}
    \centering
  \includegraphics[scale=0.8]{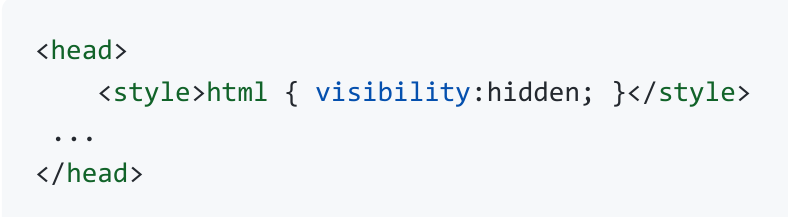}
  \Description{An html snippet is shown that shows how the website is disabled and first shown. It can be found on the GitHub site.}
   \label{fig:html2}
\end{figure}

To assure that the user can only see the webpage when it is fully loaded and cannot interact with it beforehand, the code in Figure \ref{fig:html2} can optionally be added. QButterfly will make the page visible only after all static elements are loaded.

To analyze the recorded event data of a particular user or between users, the survey results can be exported from the survey platform (e.g., csv, xlsx). QButterfly provides an Excel template (“qbutterfly.xlsx”). This contains a set of “visual basic for application” (VBA) macros for data analysis. The implemented functions “countEvent”, “countEventPattern”, and “timestamp” shown in Table \ref{tab:excel} allow us to identify specific events or event patterns and calculate intervals between events. The functions are based on regular expressions and can be combined with higher-level functions. For example, given recorded event data in the Excel cell A1, the function “=abs(timestamp(A1, “MyLink2”, 1) - timestamp(A1, “MyLink1”, 1))” would return the interval (in ms) between the user’s first click on MyLink1 and the first click on MyLink2.

%

\section{Special Considerations for Online Use}
As described in the section “Introducing QButterfly,” the overall architecture is based on the survey platform and the stimulus website. Therefore, the survey server and the website hosting must be available during the study. For this reason, in situations with many (concurrent) users accessing the stimulus website, availability must be guaranteed by an appropriate configuration. If necessary, performance testing must be carried out in advance.

Allowing participants to complete studies on their own devices, rather than in a controlled lab setting, poses several challenges for collecting reliable and precise tracking information (e.g., \cite{mathur_open-source_2019}). For example, a user’s browser in the field could theoretically render the stimulus material differently than the researcher expects because the user has a magnification function enabled or the screen resolution is too small to display the complete website. Different user devices with various performances could also lead to different timing information. Like the survey platform, QButterfly requires JavaScript to be enabled in the browser. However, users can disable JavaScript in their browsers or block specific JavaScript codes, such as cross-site JavaScript or iframes used by QButterfly. In the context of online user studies, the most efficient way to mitigate these potential problems is to conduct pre-studies to assess the target population (e.g., typical screen dimensions, active blocker extensions, and visual impairments).

\section{Validation of the Software}
\label{validation}
We validated the software in two studies – a laboratory study and a field study with crowdsourced subjects. The focus of the laboratory study was to evaluate whether the tracking information was valid and reliable. The field study investigated whether data collection is reliable in a natural environment with a broad range of devices, browsers, and individual settings. Before the two studies, the general compatibility of QButterfly with different devices and browser combinations was tested. These were Windows 10 (Chrome, Edge, Firefox, Internet Explorer), macOS (Chrome, Safari), Android (Chrome), and iOS (Chrome, Safari). Except for Internet Explorer, it worked on all platforms (see “Limitations”).
\subsection{Laboratory Study}
\subsubsection{Design}
The goal of the laboratory study was to evaluate whether a series of actions generated by a user is tracked correctly in terms of the type and sequence of the recorded events and their timestamps. Regarding the time information, an “adequate” accuracy is sufficient as compared to to high and precise temporal resolution that would be required in psychophysics experiments, for instance (e.g., \cite{anwyl-irvine_realistic_2021}). As an indication, web analytics tools, such as Google Analytics, typically report timing information in seconds instead of milliseconds \cite{google_google_nodate}. An additional concern is that the time a webpage takes to load is influenced by many factors, such as the hosting service, the network connection to and from the server, the complexity of the website (e.g., size, images, third-party content), and the performance of the client and the browser itself\footnote{Concerning images, these problems can be addressed by pre-loading the images before display \cite{garaizar_best_2019}}. These factors can lead to page load time variances in the range of seconds (e.g., \cite{kelton_modeling_2020}). A website’s content can also be dynamically loaded even after it has been successfully loaded for the first time. Therefore, the interpretation of timing information is highly dependent on the specific context.

We designed a simple study for the laboratory setting based on the “QButterfly template” for Qualtrics that recorded the user interaction. We chose a more straightforward approach instead of using external devices to register stimuli and trigger events (e.g., a BBTK photodiode and robotic actuator \cite{bridges_timing_2020}) since we considered the timing precision sufficient for our purposes. We hosted a realistic stimulus with multiple linked webpages based on HTML, CSS, and JavaScript (a fitness tracking website) on Amazon Web Services and simulated user interaction directly on the computer that showed the stimulus. Two configurations based on the most popular browsers, Chrome (95.0, Windows 10, Dell Latitude 1.9GHz i7, 16GB Ram) and Safari (15.0, macOS Catalina, MacBook Air 1.8 GHz i5, 8GB Ram) with default browser settings were selected. The interaction pattern for the website consisted of six clicks over 27 seconds. 

We developed a software bot with the Python programming language to simulate user interaction using the “PyAutoGUI” package \cite{pyautogui_welcome_nodate}. The package enables programmatic access to the mouse and keyboard. Following its development, we ran the interaction pattern a thousand times in both browsers, i.e., 2000 recorded user patterns. For each individual user-initiated event within a pattern, we calculated the delay between a click of the bot and its recognition by QButterfly (“click time delay”). To get the relevant click timepoint within the bot, we calculated the arithmetic mean of the timepoint before and directly after the execution of the Python function to generate the click. As a proxy for dwell time, we further calculated the intervals between two consecutive clicks that took the user from one webpage to the next. Finally, we compared the intervals between the bot and QButterfly (“dwell time delay”).

\subsubsection{Results}
\begin{figure}[t]
  \includegraphics[width=1\columnwidth]{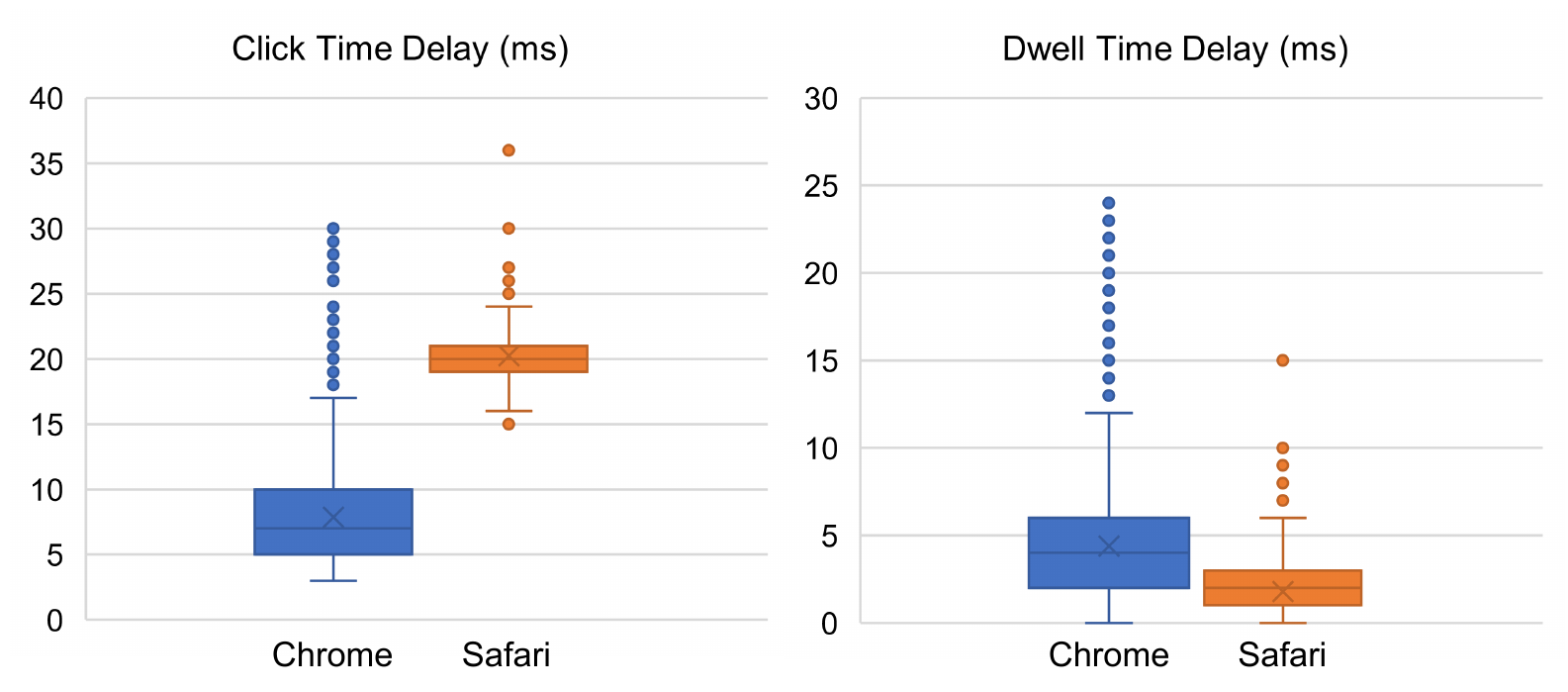}
  \caption{Box plots of the delays (ms) between the automatic click using the Python bot and the corresponding QButterfly timestamp. The click time delay (left) was calculated based on the click events, and the dwell time delay (right) was calculated based on the time between consecutive click events.}
  \label{fig:delays}
  \Description{The plot on the left shows the click time delay in ms Chrome (Mean 7) and Safari (Mean 20). The plot on the right shows the dwell time delay for Chrome (Mean 4) and Safari (Mean 2)}
\end{figure}

QButterfly successfully recorded all user actions generated by the bot, i.e., no event was missed, and no pattern was recorded in another sequence than generated. Table \ref{tab:delays} and Figure \ref{fig:delays} show the results for the two browsers. The mean delays were generally below 21ms, with only a few outliers. Only one outlier with more than four standard deviations from the mean (15 milliseconds) was detected for the dwell time delay on Safari. The QButterfly recordings were precise, with standard deviations between 1.43 and 3.63 milliseconds. For the intended purpose of enabling online user interaction studies with timings in the range of seconds, this degree of temporal precision appears to be more than sufficient.

\begin{table}[t]
  \caption{Summary of the delays (ms) between the automatic click generated by the bot and the corresponding QButterfly timestamp. The click time delay was calculated based on the click events, and the dwell time delay was based on the time between consecutive click events that brings the user from one page to the other.}
  \label{tab:delays}
 \resizebox{1\columnwidth}{!}{%
  \begin{tabular}{lcccccc}
    \toprule
    &\multicolumn{3}{c}{Click Time Delay (ms)} & \multicolumn{3}{c}{Dwell Time Delay (ms)}\\
    
   Browser&m&SD&95\% CI&m&SD&95\% CI \\
    \midrule
   Chrome (Win)&	7.88&	3.63&	[7.66, 8.10]&	4.39&	3.46&	[4.17, 4.60] \\
   Safari (Mac)&	20.23&	1.62&	[20.13, 20.33]&	1.80&	1.43&	[1.71, 1.89] \\
   \bottomrule
  \end{tabular} }
\end{table}
\subsection{Field Study}
\subsubsection{Design and Subjects}
We performed a privacy-related HCI experiment to monitor user behavior when confronted with differently designed cookie banners on websites. The purpose of cookie banners is to inform users about the meaning of tracking and collect consent. Minor design changes can significantly affect the decision to accept or decline cookies \cite{utz_informed_2019}. The IRB of our university has confirmed that no ethical approval was required, and informed consent was obtained from all individual participants. This study was limited to users with desktop computers, i.e., subjects with tablets and smartphones were excluded because the website was designed specifically for desktop screens. Users had to be UK residents, were recruited from the online panel Prolific, and were paid for their participation. We employed the “Qualtrics template” described earlier. After consent and screening, we randomly assigned users to different experimental conditions with other stimuli. Each website presented a different type of cookie banner, and users could interact with the banner and the website after accepting or declining the use of cookies. We recorded page load events as well as user clicks. At the end of the study, participants had to pass an attention check. In total, \mbox{n = 6,045} users completed the experiment, and each user’s interaction pattern was analyzed. Subjects took a median time of 4.2 minutes to complete the study.
\subsubsection{Results}
Table \ref{tab:fieldstudy} shows the characteristics of the subjects’ computers. As these were restricted to desktop computers only, operating systems, such as Android or iOS, are not represented. Subjects used a median browser window width of 1440 px (25th percentile: 1366 px; 75th percentile: 1680 px) and height of 864 px (25th percentile: 768 px; 75th percentile: 1050 px).

\begin{table}[t]
  \caption{Computing system characteristics in the field study (\mbox{n = 6,045})}
  \label{tab:fieldstudy}

  \begin{tabular}{lcclc}

  \hline
  \multicolumn{2}{c}{Browser} && \multicolumn{2}{c}{OS}\\
  \cline{1-2}\cline{4-5}
    Chrome&	71.1\% &&Windows&	73.3\% \\
Firefox&	10.0\% &&Mac	&24.5\% \\
Safari&	9.3\% &&Linux&	1.6\% \\
Edge&	8.5\% &&ChromeOS&	0.6\% \\
Opera&	1.1\% &&Others	&0.1\% \\
   \hline
   \end{tabular}
\end{table}

QButterfly worked reliably across the browser types and operating systems, and of the 6,045 recorded user interaction patterns, only a few showed any idiosyncrasy (1.3\%). Concretely, nineteen respondents had no events recorded, and 59 recordings had invalid time stamps for events (“undefined”). One of the affected users provided qualitative feedback and mentioned using a blocking tool for cross-site scripts\footnote{Cross-site scripting is a form of attack where malicious scripts are injected in an otherwise benign and trusted website (OWASP, n.d.). In the case of the QButterfly system architecture, Qualtrics and the stimulus website have different internet addresses. That is why the QButterfly JavaScript library embedded in the HTML website can be misinterpreted as a malign cross-site script and blocked if the user has a blocker installed.}. This could explain why these idiosyncrasies occurred in the first place. The remaining 5,967 recordings were further checked for plausibility (e.g., chronological order of events, missing events). 
We discovered 39 (0.6\%) interaction patterns with implausible event orders and ascertained that these particular user-click events had not been recorded correctly. This could subsequently be traced back to a mistake in embedding the QButterfly JavaScript library into the related web page’s HTML code. As a result, we introduced a function in the JavaScript library to indicate an incorrect embedding and updated the documentation on the QButterfly website. Such experiences underline the general importance of conducting pre-studies before using the library in live studies to ensure full functionality in the field.

\section{Discussion}
\subsection{Contribution}
QButterfly contributes to HCI research in the form of a validated, ready-to-use toolkit and its open-source code (cp. \cite{fogarty2017code, ledo2018evaluation}). It has been validated via a demonstration (\href{https://qbutterfly.github.io/}{qbutterfly.github.io}), usage in two HCI-related studies (\cite{ebert_bolder_2021,ebert2023saliency}), and technical measurement (Section \ref{validation}). The toolkit addresses several technical challenges researchers have when they want to conduct controlled online user interaction studies at scale \cite{hirth2017crowdsourcing}:
\begin{enumerate}
\item The software may reduce authoring time and complexity because it solves one major technical challenge, namely the disintegration of required tools \cite{hirth2017crowdsourcing}. Therefore, it is easier to design and evaluate new interactive systems at scale with large, crowdsourced participant groups, which is a demand of many HCI researchers \cite{egelman2014crowdsourcing, findlater2017}. Researchers save time in developing their study infrastructure and the data analysis after their study. 

\item It may empower non-tech-savvy researchers to conduct user interaction studies with limited programming skills without being overwhelmed by existing toolkits (e.g., psiTurk \cite{gureckis2016psiturk}). Also, behavioral researchers from other disciplines (e.g., online marketing) might be enabled to conduct online studies that would usually require rather sophisticated technical skills  \cite{hirth2017crowdsourcing}. 

\item QButterfly may facilitate the re-use of existing functionality as it is aligned with popular survey tools HCI researchers use (i.e., Qualtrics, Limesurvey) and provides easy access to a rich set of existing and evaluated functionalities and knowledge (e.g., LimeSurvey MTurk integration). Survey and experimental features such as scales, quotas, survey flow, or random assignment of participants are available “off the shelf” and "via click" in existing survey tools. Also, very complex experiments with many conditions can be carried out more quickly if these standard features can be used.

\item The toolkit may facilitate the replication of ideas by providing access to a verified study infrastructure, which enables the comparison among studies. 
\end{enumerate}
Scientific research has several advantages when researchers can conduct studies more quickly and easily. Hypotheses and artifacts can be investigated iteratively and at an earlier stage, which helps to leave wrong paths sooner. In addition, less technology-savvy researchers from other disciplines (e.g., marketing, consumer behavior, psychology) are more likely to participate in discussions if the same tools are available.
\subsection{Limitations and Future Work}
The toolkit faces several challenges typical for online studies where users use their own devices. Occasional idiosyncrasies (e.g., poor internet connection or specific browser extensions blocking events) can cause losses of tracking data, but our Excel functions automatically identify these. However, our validation study (see “Validation of the Software”) suggested that these issues only affect a small fraction of trials. While we did not observe these exceptions in our laboratory setup, about one percent of users and their tracking data were affected when data were collected in an uncontrolled, crowdsourcing setting. In a conservative approach, these individuals can be excluded from the study.
Although our implementation appears to perform reliably well across different browsers, it is incompatible with Internet Explorer. In this case, users can proceed through the questionnaire, but their tracking data are not recorded. However, the worldwide market share of Internet Explorer is steadily declining and lies at about 2 percent in 2021 \cite{kinsta_global_nodate}. 

The default instantiation of QButterfly is highly dependent on the survey platform, and the user influence on the future development of the survey software is minimal. Therefore, changes in the survey software could cause QButterfly to stop working and require adjustments to the code. For example, incompatibility issues led to the discontinuation of QRTEngine – software for running online reaction time experiments using Qualtrics \cite{barnhoorn_qrtengine_2015, van_steenbergen_qrtengine_nodate}. We have considered this by making the integration of QButterfly into the survey platform extremely lightweight. Instead of a dedicated, tightly integrated program library, very little JavaScript code is included in the survey template.

In studies using a user’s web browser, users can interact with the displayed content and other browser elements. In the case of QButterfly, users can press both the backward/forward navigation button and the browser’s refresh button. Web developers cannot fully mitigate these user actions and create additional “ready” and “load” recorded events for the affected stimulus web pages. However, QButterfly detects when a user presses the backward/forward navigation buttons and immediately redirects the user to the current page.

Future work could extend the features provided by QButterfly in its current form. While it was developed with Qualtrics in mind, the code could be readily adapted to other online survey platforms or custom interfaces, provided they support custom JavaScript code and event handling as well as the display of iframes. Furthermore, the functions that handle the analysis of the tracking data can be ported to a different environment (e.g., R, Python). Currently, they are implemented in the proprietary software MS Excel in combination with a VBA script. However, they primarily execute regular expressions to analyze events (e.g., the expression “[0-9]+\#[\textasciicircum{};]+;” matches an event), which are easy to port to different environments. 

The package currently only records user-generated keyboard or mouse clicks because the QButterfly JavaScript library embedded in the website only listens to these inputs. However, it can be extended to listen to other browser event types and record those as well (e.g., mouse trajectories generated by the “mousemove” browser event). The toolkit can also be extended for user cases requiring higher resolution timing. For example, best practices for stimulus presentation on the web, such as preloading assets, can be integrated \cite{garaizar_best_2019}. 

\subsection{Conclusion}
We present an open-source HCI toolkit (\href{http://github.com/QButterfly}{github.com/QButterfly}) that facilitates controlled online user interaction studies using the widespread Qualtrics and LimeSurvey platforms. It helps to overcome technical challenges that non-tech-savvy researchers typically face, such as the disintegration of survey tools and stimulus material (e.g., a website) and reduces the authoring time for studies. Scientists can rely on the existing functionality of their survey tools and only need to add a JavaScript library to their stimulus website to track a user’s clicks throughout their website. We empirically validated the software in a laboratory and a field study and have used it in two HCI studies. 
\begin{acks}
  We would like to thank the Hasler Foundation and the DIZH initiative of the Canton of Zurich for their financial support.
\end{acks}

\bibliographystyle{ACM-Reference-Format}
\bibliography{main}

\end{document}